\documentclass[conference]{IEEEtran}

\usepackage[scaled]{beramono}
\usepackage[T1]{fontenc}

\usepackage[square,numbers,sort&compress]{natbib}
\usepackage{xspace}
\usepackage[usenames,dvipsnames,svgnames,table]{xcolor}

\usepackage{amsmath, amssymb}

\usepackage[pdftex]{graphicx}
\usepackage{tikz}
\usetikzlibrary{arrows.meta, decorations.pathmorphing, knots, matrix, positioning, shapes, shapes.misc}
\usepackage{forest}
\usepackage{pgfplots, pgffor}

\usepackage{tcolorbox}
\usepackage[hang]{footmisc}
\usepackage{manyfoot}
\usepackage[inline]{enumitem}
\usepackage[skip=1pt,belowskip=-0.75\baselineskip,font={footnotesize}]{caption,subcaption}

\usepackage{algorithm}
\usepackage[noend]{algpseudocode}
\usepackage{listings}

\usepackage[draft,nomargin,inline,author=]{fixme}

\usepackage[colorlinks,
            pdfencoding=auto,
            bookmarksnumbered=true,
            urlcolor={blue!85!white},
            citecolor={blue!80!black},
            linkcolor={purple!80!black}]{hyperref}
\usepackage[nameinlink,capitalize,sort&compress]{cleveref} 
\graphicspath{include/imgs/}
\DeclareGraphicsExtensions{.pdf,.jpeg,.png}
\newlist{inlist}{enumerate*}{1}
\setlist*[inlist]{label=\textbf{(\arabic*)}}
\newlist{andlist}{enumerate*}{1}
\setlist*[andlist]{label=\textbf{(\arabic*)},itemjoin={{, }}, itemjoin*={{, and }}}
\newlist{orlist}{enumerate*}{1}
\setlist*[orlist]{label=\textbf{(\arabic*)},itemjoin={{, }}, itemjoin*={{, or }}}

\colorlet{success}{green!42!black}
\colorlet{failure}{red!64!black}
\colorlet{bgcolor}{white!88!black}
\colorlet{bordercolor}{white!72!black}
\colorlet{darkbordercolor}{white!64!black}
\colorlet{fgcolor}{white!48!black}

\arrayrulecolor{bordercolor}

\renewcommand{\footnoterule}{%
  \kern -2.5pt
  \hrule width 0.32\textwidth height 0.5pt
  \kern 1.5pt
}

\DeclareNewFootnote{Num}[arabic]
\DeclareNewFootnote{Sym}[fnsymbol]
\fxusetheme{color}

\let\oldReturn\Return
\renewcommand{\Return}{\State\oldReturn}

\algnewcommand{\LeftComment}[1]{\Statex \vspace*{2pt}\hspace*{-14pt}{\color{blue!42!black} $\blacktriangleright$ \:{\small #1}}}

\algnewcommand{\IfThenElse}[3]{
  \State \algorithmicif\ #1\ \algorithmicthen\ #2\ \algorithmicelse\ #3}
\algnewcommand{\IfThen}[2]{
  \State \algorithmicif\ #1\ \algorithmicthen\ #2}

\algrenewcommand\alglinenumber[1]{\color{darkbordercolor} \small \texttt{#1\:}}

\newenvironment{algfunction}[3]
    {\begingroup\textbf{func} \textsc{#1}($#2$)
     \par\leavevmode\hangindent=4em\hangafter=1\indent\leavevmode\hangindent=4em\hangafter=1\indent
     \textbf{Result:} #3
     \vspace*{1pt}
     \begin{algorithmic}[1]
     \algrenewcommand\algorithmicindent{1.25em}}
    {\end{algorithmic}\endgroup}

\newcommand{\resized}[3]{
  \resizebox{#1\linewidth}{!}{
    \begin{subfigure}[t]{#2\linewidth}
      #3
    \end{subfigure}
  }}
\newcommand{\algobox}[2]{
  \begin{tcolorbox}
        [boxrule=0.5pt,arc=2pt,boxsep=0pt,
         left=0.5pt,right=-10pt,top=3pt,bottom=3pt,
         colback=white,colframe=darkbordercolor]
  \resized{0.965}{#1}{#2}\end{tcolorbox}}

\lstdefinelanguage{SyGuS}{
  alsoletter={-,+,=,>,<},
  texcl=true, 
  morecomment=[l]{//},
  morestring=[b]",
  keywords=[1]{LIA, Int, Bool},
  keywords=[2]{set-logic, define-fun, check-synth},
  keywords=[3]{pre-f, post-f, trans-f},
  keywords=[4]{inv-f},
  keywords=[5]{synth-inv, inv-constraint, declare-primed-var},
  keywords=[6]{and, or, not, +, -, =, >, >=, <, <=}
}
\newcommand{\tturl}[1]{\texttt{\url{#1}}}
\newcommand{\mailtourl}[1]{\href{mailto:#1}{\nolinkurl{#1}}}

\newcommand{\NONE}{\texttt{None}}
\newcommand{\TRUE}{\texttt{true}}

\newcommand{\tool}[1]{\textsc{#1}\xspace}

\newcommand{\LoopInvGen}{\tool{LoopInvGen}}
\newcommand{\PIE}{\tool{PIE}}
\newcommand{\HEnum}{\tool{HEnum}}
\newcommand{\ICEDT}{\tool{ICE-DT}}

\newcommand{\Process}{\tool{Process}}
\newcommand{\Record}{\tool{Record}}
\newcommand{\Infer}{\tool{Infer}}
\newcommand{\BFL}{\tool{BFL}}
\newcommand{\Synth}{\tool{Synth}}
\newcommand{\Checker}{\tool{Check}}
\newcommand{\GetModel}{\tool{GetModel}}
\newcommand{\RecordStatesFrom}{\tool{RecordStatesFrom}}

\newcommand{\INV}{\textsc{Inv}\xspace}

\newcommand{\Inv}{\ensuremath{\mathcal{I}}\xspace}

\newcommand{\States}{\ensuremath{\mathcal{Z}}\xspace}

\newcommand{\type}[1]{\texttt{#1}\xspace}
\newcommand{\TString}{\type{String}}
\newcommand{\TInt}{\type{Int}}
\newcommand{\TBool}{\type{Bool}}
\newcommand{\TState}{\type{State}}
\newcommand{\TPred}[1]{\ensuremath{#1\rightarrow\TBool}}
\newcommand{\TSyGuS}{\type{SyGuS\textsubscript{INV}}}
\newcommand{\TConfig}{\type{Config}}

\newcommand{\SyGuSINVQuadruplet}{\ensuremath{\langle P, T, Q, \Delta \rangle}}

\begin{document}

\title{\LoopInvGen: A Loop Invariant Generator \\ based on Precondition Inference}

\author{\IEEEauthorblockN{Saswat Padhi}
\IEEEauthorblockA{University of California, Los Angeles\\
Email: \mailtourl{padhi@cs.ucla.edu}}
\and
\IEEEauthorblockN{Rahul Sharma}
\IEEEauthorblockA{Microsoft Research, India\\
Email: \mailtourl{rahsha@microsoft.com}}
\and
\IEEEauthorblockN{Todd Millstein}
\IEEEauthorblockA{University of California, Los Angeles\\
Email: \mailtourl{todd@cs.ucla.edu}}}

\IEEEspecialpapernotice{(Competition Contribution)}

\maketitle

\begin{abstract}
We describe the \LoopInvGen tool for generating loop invariants that can
provably guarantee correctness of a program with respect to a given specification.
\LoopInvGen is an efficient implementation of the inference technique
proposed in our earlier work on the precondition inference engine (\PIE).

In contrast to existing techniques, \LoopInvGen is not restricted to a fixed set of \emph{features} --
atomic predicates that are composed together to build complex loop invariants.
Instead, we start with no initial features, and use program synthesis techniques to grow the set on demand.
This not only enables a less onerous and more expressive approach,
but also appears to be significantly faster than the existing tools over
the SyGuS-Comp 2018 benchmarks from the \INV track.
\end{abstract}

\section{Introduction} \label{subsec:Introduction}

\noindent
Formally proving the correctness of a program with respect to a given specification,
can be largely automated when the appropriate \emph{program invariants} are available.
Yet, the problem of learning the adequate invariants in the first place, remains quite challenging.
Traditional \emph{static} approaches that reason over the program structure to deduce sufficient invariants,
are often inapplicable to real-life cases simply because the program logic is far too complex to be analyzable.
However, it is often the case that complex real-life programs have relatively simple invariants
that certify their correctness relative to properties of practical interest.
In such cases, \emph{data-driven} approaches seem to perform well.
These techniques learn a candidate invariant by examining program behavior (as opposed to structure),
and then refine it till it is sufficiently strong.

We extend the data-driven paradigm for inferring sufficient loop invariants.
Given some sets of ``good'' and ``bad'' program states,
data-driven approaches learn a candidate invariant as a boolean combination of atomic predicates (called \emph{features})
defined on states, such that it is satisfied by the good states and falsified by the bad ones.
Prior techniques are restricted to using a fixed set, or a fixed template for features.
For instance, a state-of-the-art technique, \ICEDT~\citep{Garg2016LearningIU}
requires the shape of constraints (such as octagonal) to be fixed apriori\footnoteNum{
    \ICEDT also requires specialized learners for boolean formulas, which can utilize the \emph{implication counterexamples}.}.
A fixed set of features not only limits the expressiveness, but predicting such a set,
which would be adequate for learning a sufficiently strong invariant is also quite challenging~\citep{Padhi2016DatadrivenPI}.

We present \LoopInvGen, a data-driven tool for inferring sufficient loop invariants,
which starts with no initial features, and automatically learns features as necessary using \emph{program synthesis} techniques.
\LoopInvGen is an optimized implementation of the general inference technique proposed in our prior work
on data-driven precondition inference~\citep{Padhi2016DatadrivenPI}.
It reduces the problem of loop invariant inference to a series of precondition inference problems,
and alternates between two phases to converge to a sufficient invariant:
\begin{andlist}
    \item \emph{learning} a candidate invariant by solving the appropriate precondition inference problem
    \item \emph{checking} if the learned candidate is sufficient for proving correctness.
\end{andlist}
If a candidate is insufficient, a \emph{counterexample} is extracted from the checker,
and is used to guide the learning phase towards the desired invariants.


Our technique is modular, and makes no assumptions on the specific program synthesizer used for feature synthesis,
except that the language of the synthesizer must be compatible with the theorem prover employed for checking.
The synthesizer utilized by \LoopInvGen is currently restricted to expressions over the theory of
\emph{linear integer arithmetic} (LIA), which is the sole focus of the \INV track of SyGuS-Comp 2019.

\begin{figure}[!t]
\centering\begin{tikzpicture}
  [minimum height=2em, text centered, text width=4em, font={\fontsize{9.5pt}{11}\selectfont}]

  \tikzstyle{arrow} = [draw, line width=0.24mm, -{Latex[length=1.75mm,width=1.75mm]}]
  \tikzstyle{oarrow} = [draw, line width=0.24mm, -{Circle[length=1.25mm,width=1.25mm]}]

  \tikzstyle{designed_comp} = [draw, rectangle, blue!40!black, fill=blue!6, rounded corners]
  \tikzstyle{existing_comp} = [draw, rectangle, cyan!30!black, fill=cyan!6, rounded corners]

  \node [text width=4.5em, minimum height=0em] (program) {Problem $\mathcal{P}$} ;
  \node [designed_comp, below=1.5em of program] (process) {\Process} ;
  \node [designed_comp, below=1.5em of process] (record) {\Record} ;
  
  \node [designed_comp, text width=12em, label={[blue!40!black, shift={(1,-1.125)}]\Infer}, minimum height=8.75em, above right=-2.625em and 2.55em of record] (infer) {} ;
  \node [label={[yellow!30!black, shift={(1.4,-0.625)}]\PIE~\citep{Padhi2016DatadrivenPI}}, draw, yellow!30!black, fill=yellow!6, rectangle, rounded corners, text width=11.125em, minimum height=3.5em, above right=-2.25em and 3em of record] (pie) {} ;

  \node [existing_comp, text width=3.25em, above right=0em and 4.5em of process] (prove) {\Checker} ;
  \node [designed_comp, magenta!30!black, fill=magenta!6, text width=2em, below left=3.75em and -1.725em of prove, minimum height=1em] (bfl) {\small \BFL} ;
  \node [existing_comp, text width=4.5em, right=2.9em of bfl, minimum height=1em] (synth) {\small \HEnum~\citep{Padhi2019OverfittingSTP}} ;
  \node [right=0.5em of infer] (inv) {Sufficient\\Invariant} ;

  \path [oarrow] (program) -- (process);
  \path [arrow] (process) -- node[right=0em] {\scriptsize Simplified $\mathcal{P}$} (record);
  \path [arrow] (process) to[bend left=15] node[above left=-0.125em and -1.25em, rotate=17.5] {\scriptsize Simplified $\mathcal{P}$} (prove);
  \path [arrow] (record) to[bend left=-2] node[below left=-0.4em and -2.25em] {\scriptsize States} (pie);
  \path [oarrow, success] (prove) to[bend left=30] (inv);
  
  \path [arrow, failure] (prove) to[bend left=20] node[text width = 5em, above right=-1.5em and 0.125em] {\scriptsize Counterexample} (pie);
  \path [arrow, success] (bfl) to[bend right=12] node[below left=-1.75em and -0.64em] {\scriptsize Candidate\\[-0.6em]Invariant} (prove);
  
  \path [arrow, failure] (bfl) to[bend left=10] node[above=-0.35em] {\scriptsize Conflicts} (synth);
  \path [arrow] (synth) to[bend left=6] node[below=-0.5em] {\scriptsize Features} (bfl);
\end{tikzpicture}
\caption{The key components in \LoopInvGen and their interdependence.}
\label{fig:overview}
\end{figure}

\section{Overview} \label{subsec:Overview}

\noindent
\Cref{fig:overview} shows a high-level schematic of \LoopInvGen.
It consists of three major components:
\begin{andlist}
    \item \Process\ -- performs simplifications using static analysis
    \item \Record\ -- collects the data required to drive the inference
    \item \Infer\ -- uses the \PIE and \Checker subcomponents to learn candidate invariants,
          and verify that they satisfy the desired properties.
\end{andlist}
The \PIE subcomponent further depends on a boolean-function learner \BFL and a feature synthesizer \Synth,
as detailed in a prior work~\citep{Padhi2016DatadrivenPI}.
We use a standard PAC learning algorithm for \BFL and
the hybrid enumeration algorithm (\HEnum)~\citep{Padhi2019OverfittingSTP} for \Synth.

In the following subsections, we briefly describe each of these subcomponents,
and illustrate them with the help of a running example.
We consider a program, listed in \cref{code:sygus},
in which $x$ is iteratively doubled starting from $1$ till $(x \geqslant y)$,
and $y$ may be arbitrarily updated at each iteration.
The goal is to verify that $(x \geqslant 1)$ always holds after the loop.
The SyGuS-\INV format~\citep{Alur2016SyGuSComp2R} used in \cref{code:sygus},
allows encoding the semantics of the program along with a desired functional specification.
For the rest of the paper, we use the quadruplet \SyGuSINVQuadruplet\ to denote an arbitrary SyGuS-\INV problem ---
$P$ being the precondition, $Q$ the postcondition, $T$ the state transition relation, and $\Delta$ the remaining relations, if any.

\begin{figure}[!t]
    \begin{tcolorbox}
          [boxrule=0.5pt,arc=2pt,boxsep=0pt,
           left=2pt,right=-10pt,top=-2pt,bottom=-3pt,
           colback=white,colframe=darkbordercolor]
\begin{lstlisting}[language=SyGuS]
(set-logic LIA)

(synth-inv inv-f ((x Int) (y Int)
                  (z1 Int) (z2 Int) (z3 Int)))

(declare-primed-var x Int)
(declare-primed-var y Int)
(declare-primed-var z1 Int)
(declare-primed-var z2 Int)
(declare-primed-var z3 Int)

(define-fun pre-f ((x Int) (y Int)
                   (z1 Int) (z2 Int) (z3 Int)) Bool
  (= x 1))


(define-fun trans-f ((x Int) (y Int)
                     (z1 Int) (z2 Int) (z3 Int)
                     (x! Int) (y! Int)
                     (z1! Int) (z2! Int) (z3! Int)) Bool
  (and (< x y) (= x! (+ x x))))

(define-fun post-f ((x Int) (y Int) 
                    (z1 Int) (z2 Int) (z3 Int)) Bool
  (or (not (>= x y)) (>= x 1)))

(inv-constraint inv-f pre-f trans-f post-f)

(check-synth)
\end{lstlisting}
    \end{tcolorbox}
    \caption{The \texttt{trex1\_vars} benchmark from SyGuS-Comp 2016 (\INV track).}
    \label{code:sygus}
\end{figure}

\subsection{\Process: Simplification using Static Analysis} \label{subsec:Process}

\noindent
This first component statically analyzes a given SyGuS-\INV problem,
and generates a simplified problem which is propagated to the subsequent components.
Moreover, it also performs basic syntactic and semantic checks to ensure validity of the problem,
and serializes it to a binary format that can be directly deserialized,
eliminating the need to re-parse the specification within the subsequent components.

Currently, the \Process component only performs an
\emph{unused variable elimination} over a given SyGuS-\INV problem \SyGuSINVQuadruplet.
For this analysis, we define ``use'' of a variable $v$ as its presence within either the specification ($P$ or $Q$),
or the state transition relation ($v$ or $v'$ in $T$), upon inlining all other relations from $\Delta$.
This analysis reduces the variables that we consider during invariant synthesis later --
unused program variables should not affect the validity of the postcondition.

To eliminate unused variables, we first construct a call graph of all the relations,
and perform 3 topological sorts over them rooted at $P$, $Q$ and $T$.
Then, starting with the leaf nodes in each sorted order, we label the ``used'' formal parameters $V_R$ for each relation $R$,
referring to the labels assigned to its callees' formal parameters, at each invocation point.
Finally, we compute the set of all used variables as:
$$
    V = V_P \cup \{ v \mid v \in V_T \vee v' \in V_T \} \cup V_Q
$$

For our running example from \cref{code:sygus}, we would have:
$$
    V_P = \{ x \} \enskip;\enskip V_T = \{ x, y, x' \} \enskip;\enskip V_Q = \{ x, y \} \qquad V = \{ x, y \}
$$

\subsection{\Record: Sampling Reachable Program States} \label{subsec:Record}

\noindent
This component collects a sample of the program states reachable at the two locations where a loop invariant must hold --
\begin{andlist}
    \item the beginning of each loop iteration
    \item just after exiting the loop.
\end{andlist}
To collect these states for programs encoded in the SyGuS-\INV format~\citep{Alur2016SyGuSComp2R},
we use a constraint solver as an execution engine\footnoteNum{
    Our original technique~\citep{Padhi2016DatadrivenPI} instrumented C/C++ programs,
    and collected program states during execution of the program.}.
We present an outline of the \Record algorithm in \cref{algo:record},
which invokes a constraint solver within the \GetModel procedure.
The algorithm accepts a SyGuS-\INV problem \SyGuSINVQuadruplet,
the set $V$ of variables to track, the desired number $n$ of program states,
and returns the set \States of the states of the variables in $V$.

\begin{figure}[!t]
    \algobox{1.1}{
    \begin{algfunction}
      {\Record}
      {\SyGuSINVQuadruplet\colon \TSyGuS, V\colon \TString[\,], n\colon \TInt}
      {A collection of program states $\States\colon \TState[\,]$.}
        \State $\States \gets \{\}$
        \While{\TRUE}
            \LeftComment{\hspace*{1.125em} Start with a \emph{previously unseen} model of the precondition.}
            \State{$m \gets \Call{GetModel}{\Delta \wedge P(m) \wedge (\bigwedge_{s \, \in \, \States} m \neq s), V}$}
            \IfThen{$m = \NONE$}{\textbf{break}}
            \State{$\States \gets \States \cup \Call{\RecordStatesFrom}{m, n}$}
            \IfThen{$|\States| = n$}{\textbf{break}}
        \EndWhile
        \Return{\States}
    \end{algfunction}
    \vspace*{4pt}
    \begin{algfunction}
        {\RecordStatesFrom}
        {m\colon \TState, k\colon \TInt}
        {A sequence $\langle m, m_1, m_2, \cdots, m_l \rangle$ of states, where $l < k$.}
          \State $\States \gets \langle m \rangle$
          \While{$|\States| < k$}
              \LeftComment{\hspace*{1.125em} Make a transition, i.e. execute a single iteration of the loop.}
              \State{$m \gets \Call{GetModel}{\Delta \wedge T(m, m'), \; \{v' \mid v \in V\}}$}
              \IfThenElse{$m = \NONE$}{\textbf{break}}{$\States \gets \States \circ \langle m \rangle$}
          \EndWhile
          \Return{\States}
      \end{algfunction}}
      \caption{An outline of the \Record component of \LoopInvGen.}
    \label{algo:record}
\end{figure}

In line 3, we start with an unseen model of the precondition,
which is a state of the program at beginning of the first iteration.
For instance, $(x \mapsto 1)$ is one such model for our running example from \cref{code:sygus}, with $V = \{x, y\}$.
The \GetModel function accepts a predicate, a list of variables,
and returns a satisfying assignment for them.
Note that this is not a \emph{complete} state of the program since the variable $y$ is unbound.
In such cases, \GetModel employs a pseudo-random number generator to extend the model to a complete program state,
assigning arbitrary values to unconstrained variables.
For our running example, such program states could, for instance,
be $(x \mapsto 1 \wedge y \mapsto -3)$, or $(x = 1 \wedge y = 7)$ etc.

In lines 5\,--\,8, we execute several iterations of the loop body, and collect the program states at the loop head each time.
In the SyGuS-\INV encoding, executing a single iteration of the loop is equivalent to making a transition from the current state.
In line 6, we solve for the next program state resulting from such a transition, and save it to \States in line 7.
For our running example, the state $(x \mapsto 1 \wedge y \mapsto 7)$ will transition to $(x \mapsto 2)$,
that could be extended to $(x \mapsto 2 \wedge y \mapsto -2)$, for example.
Note that no further transitions are possible from this state,
since $2 \not< -2$ (implicit loop guard in the transition relation).

If we reach such a state from which no transitions are possible,
and the set \States of collected program states contains less than the desired number $n$ of states then, in line 3,
we start with an \emph{unseen} state (which is not already in the set \States).

\subsection{\Infer: Generating Sufficiently Strong Loop Invariants} \label{subsec:Infer}

\noindent
This component uses the program states collected by \Record to infer a loop invariant that is sufficient
for proving correctness of a given SyGuS-\INV problem \SyGuSINVQuadruplet.
We outline our \Infer algorithm in \cref{algo:infer}, which given a SyGuS-\INV problem \SyGuSINVQuadruplet,
a set \States of reachable states, and a set $\Theta$ of configuration parameters,
returns an invariant \Inv.

\begin{figure}[!t]
    \algobox{1.1}{
    \begin{algfunction}
      {\Infer}
      {\SyGuSINVQuadruplet\colon \TSyGuS, \mathcal{Z}\colon \TState[\,], \Theta\colon \TConfig}
      {A sufficient loop invariant $\Inv\colon \TPred{\TState}$.}
        \LeftComment{Start with the weakest invariant that satisfies $\forall s\colon \Inv(s) \Rightarrow Q(s)$.}
        \State $\Inv \gets Q$
        \LeftComment{Iteratively strengthen \Inv till it is inductive.}
        \While{\TRUE}
            \State{$B \gets \{\}$}
            \While{\TRUE}
                \State{$\rho \gets \Call{\PIE}{\mathcal{Z}, B, \Theta}$}
                \State{$c \gets \Call{\Checker}{\forall s,t\colon \rho(s) \; \Rightarrow \; \Inv(s) \wedge T(s,t)\!\Rightarrow\!\Inv(t)}$}
                \IfThenElse{$c = \NONE$}{\textbf{break}}{$B \gets B \cup \{c\}$}
            \EndWhile
            \State{$\Inv \gets \Inv \wedge \rho$}
            \LeftComment{\hspace*{1.35em}Weaken \Inv using counterexamples, if it is stronger than $P$.}
            \State $c \gets \Call{\Checker}{\forall s\colon P(s) \Rightarrow \Inv(s)}$
            \If{$c \neq \NONE$}
                \State{$\mathcal{S} \gets \Call{RecordStatesFrom}{c, \Theta[\textsf{\small NumStepsOnRestart}]}$}
                \Return{$\Call{\Infer}{\SyGuSINVQuadruplet, \mathcal{Z} \cup \mathcal{S}}$}
            \ElsIf{$\rho = \TRUE$} \textbf{break}
            \EndIf
        \EndWhile
        \Return{\Inv}
    \end{algfunction}}
    \caption{An outline of the \Infer component of \LoopInvGen.}
    \label{algo:infer}
\end{figure}

A \emph{sufficient} loop invariant \Inv must satisfy three conditions:
\begin{itemize}[leftmargin=1.25em]
    \item Weaker than precondition: \hspace*{4.75pt} $\forall s\colon P(s) \Rightarrow \Inv(s)$
    \item Inductive over loop body: \hspace*{6.5pt} $\forall s,t\colon \Inv(s) \wedge T(s,t) \Rightarrow \Inv(t)$
    \item Stronger than postcondition: $\forall s\colon \Inv(s) \Rightarrow Q(s)$
\end{itemize}
As shown in \cref{fig:overview}, \Infer relies on an off-the-shelf theorem prover \Checker for verifying these conditions,
and employs \PIE~\citep{Padhi2016DatadrivenPI} to refine candidate invariants.
In line 1, it starts with the weakest possible candidate\footnoteNum{
    Our original technique~\citep{Padhi2016DatadrivenPI} used PIE to learn the initial candidate invariant \Inv as
    one that satisfies $\{\Inv\}\;\textbf{skip}\;\{Q\}$.
    We found this initial candidate to be too strong sometimes, requiring additional counterexamples to weaken it.},
$\Inv = Q$, and iteratively refines \Inv till all of the above properties are satisfied.
For instance, on our running example from \cref{code:sygus},
\Infer starts with the initial candidate invariant $\Inv_0 = ((x < y) \vee (x \geqslant 1))$.

However, this candidate invariant is not inductive.
The state $(x \mapsto 0 \wedge y \mapsto 1)$ satisfies \Inv,
but it may transition to state $(x \mapsto 0 \wedge y \mapsto 0)$, which violates \Inv.
In lines 2\,--\,13, \Infer employs a \emph{strengthening} loop (inspired by \tool{HOLA}~\citep{Dillig2013InductiveIG}),
to ensure inductiveness of the candidate.
At the $i^\text{th}$ iteration, it learns a precondition $\rho_i$
under which the candidate invariant is preserved after a single transition.
For our running example, $\rho_1 = (x \geqslant 1)$, for instance,
would ensure that our candidate invariant $\Inv_0 = ((x < y) \vee (x \geqslant 1))$ is preserved.
In line 8, we strengthen the candidate invariant by conjoining it with the learned precondition.
For our running example, the new candidate $\Inv_1 = \Inv_0 \wedge \rho_1 = (x \geqslant 1)$ is indeed inductive.

The reduction to a precondition inference problem allows us to leverage our prior work, \PIE,
on learning preconditions with automatic synthesis of appropriate features\footnoteNum{
    \PIE uses two off-the-shelf components:
    \begin{andlist}
        \item a program synthesizer \Synth to generate new features
        \item a boolean function learner \BFL to learn a composition of these features.
    \end{andlist}
    The details are presented in our full paper~\citep{Padhi2016DatadrivenPI}.}.
In line 5, \PIE accepts a set \States of states which lead to satisfaction of a desired property,
a set $B$ of states which do not, the set $\Theta$ of configuration parameters
(such as \emph{conflict group size}~\citep{Padhi2016DatadrivenPI}),
and learns a \emph{likely} precondition $\rho$ for the desired property.
Since the precondition is only a likely one, in line 6, \Infer checks the likely precondition using \Checker for sufficiency,
and provides counterexamples to \PIE iteratively, in lines 4\,--\,7, till a provably sufficient precondition is learned.

Conjoining the current candidate invariant \Inv with the precondition $\rho$,
might however result in the next candidate $\Inv \wedge \rho$ being too strong,
in particular, stronger than the precondition $P$.
Therefore, in line 9, we use \Checker to verify that it is weaker than $P$.
A counterexample in this case would indicate a state that is allowed by the precondition,
but not covered by the candidate invariant.
This could happen due to inadequate exploration of program states during the \Record phase,
for instance due to a complex transition relation.
On finding such a counterexample $c$, we invoke \RecordStatesFrom (from \cref{algo:record})
to collect a few more (\textsf{\small NumStepsOnRestart} parameter in $\Theta$) states starting from $c$,
in line 11, to account for the unexplored program behavior.
Finally, in line 12, we restart with the new set of available program states.
Note that if no such counterexample is found and the current candidate \emph{unconditionally} holds (i.e. $\rho = \TRUE$),
as is the case with the candidate $\Inv = (x \geqslant 1)$ for our running example, then
our current candidate invariant is provably sufficient for guaranteeing correctness of \SyGuSINVQuadruplet.

\section{Implementation} \label{subsec:Implementation}

\noindent
Our implementation of \LoopInvGen is open source, and is available at {\small \tturl{https://github.com/SaswatPadhi/LoopInvGen}}.
For its various components, \LoopInvGen internally uses the following off-the-shelf algorithms or implementations:
\begin{itemize}
    \item Both \GetModel and \Checker are implemented using the \tool{Z3}~\citep{Moura2008Z3AE} theorem prover.
          Our prior work used \tool{CVC4}~\citep{Barrett2011CVC4} for reasoning over the theory of strings,
          which is beyond the scope of \INV track of SyGuS-Comp 2018.
          However, constraint solving with both \tool{Z3} and \tool{CVC4} in parallel is still in progress.
    
    \item \PIE uses the \HEnum~\citep{Padhi2019OverfittingSTP} algorithm for grammar-based expression synthesis.
          The language for synthesis has been shrunk to only allow expressions over LIA theory.
    
    \item The \BFL component in \PIE uses a standard \emph{probably approximately correct} (PAC) algorithm
          that can learn arbitrary \emph{conjunctive normal form} (CNF) formula,
          and is biased towards small formulas~\citep{Kearns1994AnIT}.
\end{itemize}

\subsection*{\textsc{Major Implementation Changes}}

\subsubsection*{Since SyGuS-Comp 2018}
\begin{itemize}
    \item \tool{PLearn}~\citep{Padhi2019OverfittingSTP} --
          We now run several parallel instances of \Infer each with a different subgrammar of the LIA/NIA grammar.
          This significantly reduces chances of overfitting and hence improves inference time,
          as described in our recent work~\cite{Padhi2019OverfittingSTP}.

    \item Hybrid Enumeration~\citep{Padhi2019OverfittingSTP} --
          Within \PIE, we now use the hybrid enumeration technique for feature synthesis.
          As detailed in our recent work~\cite{Padhi2019OverfittingSTP},
          this technique interleaves exploration of subgrammars of varying expressiveness,
          and typically results in significantly faster convergence.

    \item Early Postcondition Check --
          The \Process{} component now checks if the postcondition itself is a sufficient loop invariant,
          even before sampling any program states (\Record) and invoking inference (\Infer).

    \item Multiple Counterexamples --
          We observed that adding multiple counterexamples (32) instead of just one (see line 6 of \cref{algo:infer})
          significantly reduces the number of CEGIS rounds and results in faster convergence.
\end{itemize}

\subsubsection*{Since SyGuS-Comp 2017}
\begin{itemize}
    \item \Process{} --
          We now have a static analysis pass before the \Record and \Infer components (see \cref{subsec:Process}).

    \item Early Precondition Check --
          As opposed to finally checking if an inductive invariant is weaker than the precondition,
          we now check this property at each strengthening.

    \item AST Pruning --
          We have implemented a syntactic checking phase before \tool{Escher}'s semantic checks,
          that prunes redundant ASTs such as (\texttt{\_ + x - x}) or (\texttt{1 * \_}) etc.

    \item Better SyGuS-IF~\citep{Alur2016SyGuSComp2R} Support --
          We have added support for defining and invoking arbitrary auxiliary relations,
          other than precondition, postcondition and transition relations.

    \item Beyond LIA Theory --
          We have implemented experimental support for theory of Non-Linear Integer Arithmetic (NLIA),
          which may be activated using the command: \texttt{(set-logic NLIA)}.
\end{itemize}

\subsubsection*{Since First Publication~\citep{Padhi2016DatadrivenPI}}
\begin{itemize}
    \item \Record Coverage --
          The \Record component has been significantly improved to better explore program states for non-deterministic programs.
          Along with a better selection of initial candidate invariant,
          this allowed us to start with only 512 program states instead of 6400.

    \item Parallel \Record\ --
          Multiple (by default, 2) instances of \Record with different seeds for PRNGs are run in parallel,
          and the program states are then merged.

    \item \tool{Z3} Scopes --
          \LoopInvGen creates a single subprocess for \tool{Z3}, and relies heavily on scopes to cache context information,
          and minimize the size of queries.

    \item Unsolvability Detection --
          \LoopInvGen immediately terminates if $\exists s\colon P(s) \not\Rightarrow Q(s)$,
          i.e. the precondition does not imply the postcondition.
          It also keeps track of known program states, and terminates as soon as a state appears to be a negative example
          (w.r.t. the given specification).

    \item \emph{Conflict Group Size}~\citep{Padhi2016DatadrivenPI} --
          Overriding \PIE's default size of $16$, \LoopInvGen uses $64$.
\end{itemize}

\section{Conclusion} \label{subsec:Conclusion}

\noindent
We have described \LoopInvGen, which uses a data-driven approach to generate loop invariants
that provably guarantee the correctness of an implementation with respect to a given specification.
In contrast to existing techniques, \LoopInvGen
\begin{andlist}
    \item is not restricted to any specific logical theory
    \item starts with no initial features and learns them automatically on demand.
\end{andlist}
In essence, \LoopInvGen reduces the problem of loop invariant inference to a series of precondition inference problems,
and solves them using \PIE, which uses a form of program synthesis to learn features in a targeted manner.

\section*{Acknowledgment}

\noindent
We thank the organizers of SyGuS-Comp for making all the solvers and benchmarks publicly available.

\bibliographystyle{plainnat}
\begingroup
  \small
  \bibliography{refs}
\endgroup

\end{document}